\documentclass[12pt]{extarticle}
\usepackage{setspace}

\usepackage{setspace}
\usepackage[T1]{fontenc}
\usepackage{times}
\usepackage{palatino} 
\usepackage{lmodern}
\usepackage[latin1]{inputenc}
\usepackage{epsfig}
\usepackage[english]{babel}
\usepackage{color}
\usepackage{graphicx}
\usepackage{dcolumn}
\usepackage{moreverb}
\usepackage{amsmath,amssymb,amsfonts}
\usepackage[all]{xy}

\usepackage[svgnames]{xcolor}
\usepackage{tikz}

\def\nudge{.5}

\tikzset{axis/.style={ultra thick, Red!75!black, -latex, shorten <=-\nudge cm, shorten >=-2*\nudge cm}}
\tikzset{line/.style={thick,Green}}

\begin{document}
\numberwithin{equation}{section}
\newcommand{\boxedeqn}[1]{%
  \[\fbox{%
      \addtolength{\linewidth}{-2\fboxsep}%
      \addtolength{\linewidth}{-2\fboxrule}%
      \begin{minipage}{\linewidth}%
      \begin{equation}#1\end{equation}%
      \end{minipage}%
    }\]%
}


\newsavebox{\fmbox}
\newenvironment{fmpage}[1]
     {\begin{lrbox}{\fmbox}\begin{minipage}{#1}}
     {\end{minipage}\end{lrbox}\fbox{\usebox{\fmbox}}}

\raggedbottom
\onecolumn

\parindent 8pt
\parskip 10pt
\baselineskip 16pt
\noindent\title*{{\LARGE{\textbf{Recurrence approach and higher rank polynomial algebras for superintegrable monopole systems }}}}
\newline
\newline
\newline
Md Fazlul Hoque$^a$, Ian Marquette$^a$ and Yao-Zhong Zhang$^{a,b}$
\newline
\newline
$a.$ School of Mathematics and Physics, The University of Queensland, Brisbane, QLD 4072, Australia
\newline
\newline
$b.$ Institute of Theoretical Physics, Chinese Academy of Sciences, Beijing 100190, China
\newline
\newline
E-mail: m.hoque@uq.edu.au; i.marquette@uq.edu.au; yzz@maths.uq.edu.au
\newline
\newline
\begin{abstract}
We revisit the MIC-harmonic oscillator in flat space with monopole interaction and derive the polynomial algebra satisfied by the integrals of motion and its energy spectrum using the ad hoc recurrence approach. We introduce a superintegrable monopole system in generalized Taub-NUT space. The Schr\"{o}dinger  equation of this model is solved in spherical coordinates in the framework of St\"{a}ckel transformation. It is shown that wave functions of the quantum system can be expressed in terms of the product of Laguerre and Jacobi polynomials. We construct ladder and shift operators based on the corresponding wave functions and obtain the recurrence formulas. By applying these recurrence relations, we construct higher order algebraically independent integrals of motion. We show the integrals form a polynomial algebra. We construct the structure functions of the polynomial algebra and obtain the degenerate energy spectra of the model. 
\end{abstract}

\section{Introduction}

The connection between quantum models and magnetic monopoles, integrable and superintegrable systems is well-known. In this paper, we show that the same connection applies to the harmonic oscillator with Abelian monopole using the recurrence approach. To our knowledge the recurrence approach had not previously been applied to Hamiltonian systems with magnetic monopole interactions. 

 In classical and quantum mechanical systems, constructive approach is one of the powerful tools to derive integrals of motion. The first- and second-order ladder operators have been used by several authors to construct integrals of motion and their corresponding polynomial algebras \cite{Jau1, Fri1, Boy1, Eva1, Mar1}. There are many distinct approaches for the constructions of integrals of motion using higher order ladder operators (see e.g. \cite{Kre1, Adl1, Jun1, Dem1, Mar2, Rag1, Mar4, Mar3, Mar5}). In fact, there is a close connection between recurrence approach and special functions and orthogonal polynomials \cite{Kal1}. The operator version of recurrence relations, their algebraic relations \cite{Cal1, Cal2} and connection to Lissajous models related to Jacobi exceptional orthogonal polynomials were investigated \cite{Mar6}. Recently the authors in the present paper applied coupling constant metamorphosis to systems amenable to the ladder operator method \cite{FH3}. However most these previous studies have been restricted to systems with scalar potentials.
 
Superintegrable systems with non-scalar potentials such as spin \cite{Win1, Nik1}, magnetic field and magnetic monopole \cite{Wu1, Jac1, Hok1, Lab1, Mad4} have recently attracted much interest. In \cite{Mar7}, a quantum superintegrable system in the field of Kaluza-Klein magnetic monopole was studied. We are interested superintegrable monopole system in space with Taub-NUT metric. The geodesic of the Taub-NUT metric appropriately describes the motion of well-separated monopole-monopole interactions ( see e.g. \cite{Cor1, Iwa1, Iwa6, Gro1, Cot1, Gib1}). Recently we introduced a Kepler quantum monopole system in a generalized Taub-NUT space which includes the Kaluza-Klein and MIC-Zwanziger monopole systems as special cases \cite{FH4}.

The purpose of the present paper is twofold: Firstly we revisit the MIC-harmonic oscillator in the field of magnetic monopole in flat space \cite{Lab1} by means of a somewhat ad hoc recurrence approach. We construct the integrals of motion and (polynomial) algebraic relations satisfied by them. This enables us to present an algebraic derivation of the energy spectrum of the system. Secondly we introduce a new MIC-harmonic oscillator type Hartmann system with monopole interaction in a generalized Taub-NUT space. We construct its integrals of motion using recurrence relations based on wave functions. We show the integrals satisfy a higher-order polynomial algebra and apply this algebraic structure to derive the energy spectrum.

\section{MIC-harmonic oscillator with monopole in flat space}
The problem of the accidental degeneracies in the spectrum of a harmonic oscillator in the field of magnetic monopole was investigated in \cite{Lab1, Mci1}. In this section we revisit this model using a somewhat ad hoc recurrence method.

Consider the Hamiltonian with monopole interaction 
\begin{eqnarray}
H=\frac{1}{2}\left[\textbf{p}^2+\frac{c_0 r^2}{2}+\frac{Q^2}{r^2}\right],\label{ham1}
\end{eqnarray}
where $p_i=-i\partial_i-A_i Q$; $A_1=\frac{-y}{r(r+z)}$, $A_2=\frac{x}{r(r+z)}$,  and $A_3=0$ are the 3 components of the vector potential associated with the magnetic monopole; $c_0$ and $Q$ are constants. This system is the well-known MIC-harmonic oscillator \cite{Mci1}.
Setting $Q^2=\lambda$ and $\frac{c_0}{2}=\omega^2$, the Hamiltonian becomes the one in \cite{Lab1}.
The total angular momentum of the system reads
\begin{eqnarray}
\textbf{L}=\textbf{r}\times \textbf{p}-Q\frac{\textbf{r}}{r}.
\end{eqnarray}
The eigenvalues of $\textbf{L}^2$ are, as usual, of the form $l(l+1)$ with $l=|Q|, |Q|+1, |Q|+2,\dots$. Let 
\begin{eqnarray}
T=-\frac{1}{4}(\textbf{r}.\textbf{p}-\textbf{p}.\textbf{r}),\quad
S=-\frac{1}{2\omega}\left(\frac{1}{2}\textbf{p}^2+\frac{Q^2}{2r^2}-\frac{\omega^2 r^2}{2}\right).
\end{eqnarray}
Then $T$, $S$ and $\frac{1}{2\omega}H$ satisfy the $O(2,1)$ commutation relations and eigenstates of $H$ belong to irreducible $O(2,1)$ representation spaces \cite{Lab1}. The eigenvalues of $H$ is of the form $2\omega(d^{+}_l+n)$,  where $n=0, 1, 2,\dots$ and $d^{+}_l=\frac{1}{2}\{1+ (l+\frac{1}{2})\}$. As pointed out in \cite{Lab1}, a complete set of quantum number is obtained by simultaneously diagonalizing $H$, $\textbf{L}^2$, $L_3$. The action of $H$, $\textbf{L}^2$ and $L_3$ on the basis vectors $|n,l,m\rangle$ is given by
\begin{eqnarray}
&H|n,l,m\rangle&=2\omega(d^{+}_l+n)|n,l,m\rangle,\quad n=0, 1, 2,\dots,\label{h1}
\\&
\textbf{L}^2|n,l,m\rangle&= l(l+1)|n,l,m\rangle, \quad l=|Q|, |Q|+1,\dots,\label{l1}
\\&
L_3|n,l,m\rangle&= m|n,l,m\rangle,\quad m=-l, -l+1,\dots, l-1, l. \label{l2}
\end{eqnarray}
The physical energy spectrum of $H$ is found to be 
\begin{eqnarray}
E_{k}=\omega(k+\frac{3}{2}), \quad k=2n+l.\label{pe1}
\end{eqnarray}
Introduce \cite{Lab1}
\begin{eqnarray}
a_i=\frac{1}{\sqrt{2}}(u_i+\frac{i}{\omega}\dot{u_i})\quad \text{and}\quad a^\dagger_i=\frac{1}{\sqrt{2}}(u_i+\frac{i}{\omega}\dot{u_i}),
\end{eqnarray}
where $u_i=\frac{\epsilon_{ijk}}{\sqrt{2}}(L_j r_k+r_k L_j)$ and $\dot{u_i}=\frac{\epsilon_{ijk}}{\sqrt{2}}(L_j v_k+v_k L_j)$, $i,j,k=1,2,3$. They also satisfy the commutation relations  $[H, a^\dagger_i]=\omega a^\dagger_i$ and $[H, a_i]=-\omega a_i$. Let 
\begin{eqnarray}
A=\frac{1}{2\omega}(H+\omega B-\omega), \quad H_{\pm}\equiv S\pm iT,
\end{eqnarray}
where $B=\sqrt{\textbf{L}^2+\frac{1}{4}}$ which is a well-defined operator \cite{Lab1}.

 We now construct ladder operators \begin{eqnarray}
A X^{+}=A a^\dagger_3-H_{+}a_3 \quad \text{and}\quad X^{-}A= a_3 A-a^\dagger_3 H_{-}.
\end{eqnarray}
Then on the basis vectors $|n,l,m\rangle$, 
\begin{eqnarray}
&H_{+}|n,l,m\rangle&=\sqrt{(n+1)(n+l+\frac{3}{2})} |n+1,l,m\rangle,
\\&
H_{-}|n,l,m\rangle &=\sqrt{n(n+l+\frac{1}{2})} |n-1,l,m\rangle,
\\
&a_3|n,l,m\rangle & =c_0(n, l-1, m) |n,l-1,m\rangle +c_1(n-1, l+1,m)\nonumber\\&&\quad\times |n-1,l+1,m\rangle,
\\&
a^\dagger_3|n,l,m\rangle &=c^{*}_0(n, l, m) |n,l+1,m\rangle +c^{*}_1(n, l,m)|n+1,l-1,m\rangle,\nonumber\\&&
\\
&AX^{+}|n,l,m\rangle &=(l+\frac{3}{2})c^{*}_0(n,l,m)|n,l+1,m\rangle,
\\&
X^{-}A|n,l,m\rangle &=(l+\frac{1}{2})c_0(n,l-1,m)|n,l-1,m\rangle,  
\end{eqnarray}
where
\begin{eqnarray*}
&&c_0(n,l,m)=-i\sqrt{\frac{(2n+2l+3)(l-m+1)(l+m+1)(l-Q+1)(l+Q+1)}{\omega(2l+1)(2l+3)}},
\\&&
c_1(n,l,m)=i\sqrt{\frac{2(n+1)(l-m)(l+m)(l-Q)(l+Q)}{\omega(2l-1)(2l+1)}}.
\end{eqnarray*}

\subsection{Integrals of motion, algebra structure and unirreps}
We now take the combinations 
\begin{eqnarray}
D_1 = H_{+}(X^{-}A)^2(B-2), \quad D_2=(B-2)(AX^{+})^2 H_{-} 
\end{eqnarray}
whose action on the basis vectors show the raising and lowering of quantum numbers while preserving energy $E$. We have 
\begin{eqnarray}
D_1|n,l,m\rangle &=&(l-\frac{3}{2})(l-\frac{1}{2})(l+\frac{1}{2})\sqrt{(n+1)(n+l-\frac{1}{2})} c_0(n,l-1,m)\nonumber\\&&\times c_0(n,l-2,m)|n+1,l-2,m\rangle,
\\
D_2|n,l,m\rangle &=&(l+\frac{1}{2})(l+\frac{3}{2})(l+\frac{5}{2})\sqrt{n(n+l+\frac{1}{2})} c^{*}_0(n-1,l,m)\nonumber\\&&\times c^{*}_0(n-1,l+1,m)|n-1,l+2,m\rangle.  
\end{eqnarray}
We can also obtain the action of the operators $D_1D_2$ and $D_2D_1$ on the basis vectors. Then together with (\ref{h1}), (\ref{l1}) and (\ref{l2}), we can conclude that on the operator level,
\begin{eqnarray}
&&[D_1, H]=0=[D_2, H], \quad [D_1, L_3]=0=[D_2, L_3],\label{kffh1}
\\&&
[B, D_1]= -2D_1,\quad \qquad [B, D_2]= 2D_2,\label{kffh2}
\end{eqnarray}
\begin{eqnarray}
&D_1D_2&=\frac{B(B+2)}{16384\omega^6 }[2B-2L_3-1][2B-2L_3+1][2B+2L_3-1]\nonumber\\&&\quad\times [2B+2L_3+1][2B-2Q-1][2B-2Q+1][2B+2Q-1]\nonumber\\&&\quad\times[2B+2Q+1][H+\omega B-\omega]^2[H-\omega B-\omega][H+\omega B+\omega],\nonumber\\&&\quad\label{kffh3}
\end{eqnarray}
\begin{eqnarray}
&D_2D_1&=\frac{(B-2)B}{16384\omega^6 }[2B-2L_3-3][2B-2L_3-1][2B+2L_3-3]\nonumber\\&&\quad\times [2B+2L_3-1][2B-2Q-3][2B-2Q-1][2B+2Q-3]\nonumber\\&&\quad\times[2B+2Q-1][H+\omega B-3\omega]^2[H-\omega B+\omega][H+\omega B-\omega].\nonumber\\&&\quad\label{kffh4}
\end{eqnarray}
Thus $D_1$, $D_2$ and $B$ form a higher-order polynomial algebra with central elements $H$ and $L_3$.
In order to derive the spectrum using the polynomial algebra, we realize this algebra in terms of deformed oscillator algebra \cite{Das2, Das1} $\{\aleph, b^{\dagger}, b\}$ of the form
\begin{eqnarray}
[\aleph,b^{\dagger}]=b^{\dagger},\quad [\aleph,b]=-b,\quad bb^{\dagger}=\Phi (\aleph+1),\quad b^{\dagger} b=\Phi(\aleph).\label{kpfh}
\end{eqnarray}
Here $\aleph $ is the number operator and $\Phi(x)$ is well behaved real function satisfying 
\begin{eqnarray}
\Phi(0)=0, \quad \Phi(x)>0, \quad \forall x>0.\label{kpbc}
\end{eqnarray}
We rewrite ((\ref{kffh1})-(\ref{kffh4})) in the form of deformed oscillator (\ref{kpfh}) by letting $\aleph=\frac{B}{2}$, $b=D_1$ and $b^{\dagger}=D_2$. We then obtain the structure function
\begin{eqnarray}
&\Phi(x;u,E)&=\frac{(2x+u)(2x+u-2)}{16384\omega^6}[E+\omega(2x+u-3)]^2[E-\omega(2x+u-1)]\nonumber\\&&\times [E+\omega(2x+u-1)][2(u+2x)-2L_3-3][2(u+2x)-2L_3-1]\nonumber\\&&\times [2(u+2x)+2L_3-3][2(u+2x)+2L_3-1][2(u+2x)-2Q-3]\nonumber\\&&\times [2(u+2x)-2Q-1][2(u+2x)+2Q-3][2(u+2x)+2Q-1],\nonumber\\&&
\end{eqnarray}
where $u$ is arbitrary constant. In order to obtain the $(p+1)$-dimensional unirreps, we should impose the following constraints on the structure function
\begin{equation}
\Phi(p+1; u,E)=0,\quad \Phi(0;u,E)=0,\quad \Phi(x)>0,\quad \forall x>0,\label{pro2}
\end{equation}
where $p$ is a positive integer. These constraints give $(p+1)$-dimensional unitary representations and their solution gives the energy $E$ and the arbitrary constant $u$. We have the following possible constant $u$ and energy spectra $E$, for the constraints $\varepsilon_1=\pm 1$, $\varepsilon_2=\pm 1$, $\varepsilon_3=\pm 1$:
\begin{eqnarray}
&&u=\frac{1}{2}(1+2\varepsilon_1 m), \quad E=\frac{\varepsilon_2 \omega}{2}[2+\varepsilon_3(1+4p)+2\varepsilon_1 m];\label{e1}
\\&&
u=\frac{1}{2}(1+2\varepsilon_1 Q), \quad E=\frac{\varepsilon_2 \omega}{2}[2+\varepsilon_3(1+4p)+2\varepsilon_1 Q] ;
\\&&
u=\frac{1}{\omega}(\omega+\varepsilon_1 E), \quad E=\frac{\varepsilon_2 \omega}{2}(3+2p+2\varepsilon_1 m);\label{e2}
\\&&
u=\frac{1}{\omega}(\omega+\varepsilon_1 E), \quad E=\frac{\varepsilon_2 \omega}{2}(3+2p+2\varepsilon_1 Q).
\end{eqnarray}
Making the identification $p=n$, $l=m$, $\varepsilon_1=1$, $\varepsilon_2=1$, $\varepsilon_3=1$, the energy spectra (\ref{e1}) and (\ref{e2}) coincide with the physical spectra (\ref{pe1}). The physical wave functions involve other quantum numbers and we have in fact the degeneracy of $p$ only when these other quantum numbers would be fixed.

\section{MIC-harmonic oscillator with monopole in generalized Taub-NUT space}
Let us consider the generalized Taub-NUT metric in $\mathbb{R}^3$ 
\begin{eqnarray}
ds^2=f(r)dl^2+g(r)(d\psi+A_i d\textbf{r})^2,\label{mc1}
\end{eqnarray}
where 
\begin{eqnarray}
&&f(r)=a r^2+b, \quad g(r)=\frac{r^2(ar^2+b)}{1+c_1 r^2+d r^4},\label{fg1}
\\&&
A_1=\frac{-y}{r(r+z)}, \quad A_2=\frac{x}{r(r+z)}, \quad A_3=0,
\end{eqnarray}
$r=\sqrt{x^2+y^2+z^2}$ and the three dimensional Euclidean line element $dl^2=dx^2+dy^2+dz^3$, $a$, $b$, $c_1$, $d$ are constants. Here $\psi$ is the additional angular variable which describes the relative phase and its coordinate is cyclic with period $4\pi$ \cite{Cor1, Grs1}. The functions $f(r)$ and $g(r)$ in the metric represent gravitational effects and $A_i$ are components of the potential associated with the magnetic monopole field.

We consider the Hamiltonian associated with (\ref{mc1}) 
\begin{eqnarray}
H=\frac{1}{2}\left[\frac{1}{f(r)}\left\{\textbf{p}^2+\frac{c_0 r^2}{2}+c_4\right\}+\frac{Q^2}{g(r)}\right]\label{kf1},
\end{eqnarray}
where $c_0$ and $c_4$ are constants and the operators 
\begin{eqnarray}
p_i=-i(\partial_i-iA_i Q), \quad Q=-i\partial_\psi
\end{eqnarray}
satisfying the following commutation relations
\begin{eqnarray}
[p_i,p_j]=i\epsilon_{ijk}M_k Q, \quad [p_i,Q]=0,\quad \textbf{M}=\frac{\textbf{r}}{r^3}.
\end{eqnarray} 
The system with Hamiltonian (\ref{kf1}) is a Hartmann system in a curved Taub-NUT space with abelian monopole interaction. This new system is referred to as MIC-harmonic oscillator monopole system. In this section, we solve the Schr\"{o}dinger St\"{a}ckel equivalent of the system (\ref{kf1}) in spherical coordinates, derive the recurrence relations and construct higher order integrals and the corresponding higher rank polynomial algebra.

\subsection{Separation of variables}
Let us consider the spherical coordinates 
\begin{eqnarray}
&&x=r \sin\theta\cos\phi,\quad y=r\sin\theta\sin\phi,\quad z=r\cos\theta,
\end{eqnarray}
where $r>0$, $0\leq\theta\leq\pi$ and $0\leq \phi\leq 2\pi$.
In terms of these coordinates, the Taub-NUT metric (\ref{mc1}) takes on the form 
\begin{eqnarray}
ds^2=f(r)(dr^2+r^2d\theta^2+r^2\sin^2\theta d\phi^2)+g(r)(d\psi+\cos\theta d\phi)^2,\label{mc2}
\end{eqnarray}
\begin{eqnarray}
A_1=-\frac{1}{r}\tan\frac{\theta}{2}\sin\phi,\quad A_2=\frac{1}{r}\tan\frac{\theta}{2}\cos\phi,\quad A_3=0.
\end{eqnarray}
The Schr\"{o}dinger equation of the model (\ref{kf1}) is
\begin{eqnarray} 
&H\Psi(r,\theta,\phi,\psi)&=\frac{-1}{2(a r^2+b)}\left[\frac{\partial^2}{\partial r^2}+\frac{2}{r}\frac{\partial}{\partial r}-\frac{c_0 r^2}{2}-c_4\right.\nonumber\\&&\left.+\frac{1}{r^2}\left(\frac{\partial^2}{\partial\theta^2}+\cot\theta\frac{\partial}{\partial\theta}+\frac{1}{\sin^2\theta}\frac{\partial^2}{\partial\phi^2}\right)+\left(\frac{1}{r^2\cos^2\frac{\theta}{2}}+c_1+d r^2\right)\right.\nonumber\\&&\left.\times\frac{\partial^2}{\partial\psi^2}-\frac{1}{r^2\cos^2\frac{\theta}{2}}\frac{\partial}{\partial\phi}\frac{\partial}{\partial\psi}\right] \Psi(r,\theta,\phi,\psi) =E\Psi(r,\theta,\phi,\psi).\label{kf2}
\end{eqnarray}
We can write $\Psi(r,\theta,\phi,\psi)=\chi(r,\theta)e^{i(\nu_1\phi+\nu_2\psi)}$. Then we obtain the equivalent system of (\ref{kf2}) as
\begin{eqnarray} 
&H'\chi(r,\theta)e^{i(\nu_1\phi+\nu_2\psi)}&=\left[\frac{\partial^2}{\partial r^2}+\frac{2}{r}\frac{\partial}{\partial r}+\left(2aE-d\nu_2^2-\frac{c_0 }{2}\right)r^2\right.\nonumber\\&&\left.+\frac{1}{r^2}\left(\frac{\partial^2}{\partial\theta^2}+\cot\theta\frac{\partial}{\partial\theta}-\frac{\nu_1^2}{\sin^2\theta}\right)-\frac{\nu_2^2}{r^2\cos^2\frac{\theta}{2}}+\frac{\nu_1\nu_2}{r^2\cos^2\frac{\theta}{2}}\right]\nonumber\\&&\times\chi(r,\theta)e^{i(\nu_1\phi+\nu_2\psi)} =E'\chi(r,\theta)e^{i(\nu_1\phi+\nu_2\psi)},\label{kff2}
\end{eqnarray}
where $E'=c_4+c_1\nu_2^2-2bE$. The original energy parameter $E$ now plays as the role of model parameter and the model parameter $c_4+c_1\nu_2^2-2bE$ plays the role of energy $E'$. This change in the role of the parameters is called coupling constant metamorphosis. Moreover, the model (\ref{kff2}) is related to the one in (\ref{kf2}) by St\"{a}ckel transformation and thus the two systems are St\"{a}ckel equivalent \cite{Boy2,Kal3}.

By making the Ansatz,
\begin{eqnarray}
\Psi(r,\theta,\phi,\psi)=R(r)\Theta(\theta)e^{i(\nu_1\phi+\nu_2\psi)},
\end{eqnarray}
(\ref{kff2}) becomes the radial and angular ordinary differential equations
\begin{eqnarray}
&&\left[\frac{\partial^2}{\partial r^2}+\frac{2}{r}\frac{\partial}{\partial r}-E'+(2aE-d\nu_2^2-\frac{c_0 }{2}) r^2 -\frac{k_1}{r^2}\right ]R(r)=0,\label{kf5}
\\
&&\left[\frac{\partial^2}{\partial\theta^2}+\cot\theta\frac{\partial}{\partial\theta}+\left\{k_1-\frac{(\nu_1-2\nu_2)^2}{2(1+\cos\theta)}-\frac{\nu_1^2}{2(1-\cos\theta)}\right\}\right]\Theta(\theta)=0,\nonumber\\&&\label{kf6}
\end{eqnarray}
where $k_1$ is separable constant. 

We now turn to (\ref{kf6}), which can be converted, by setting $z=\cos\theta$ and $\Theta(z)=(1+z)^{a}(1-z)^{b} Z(z)$, to
\begin{eqnarray}
&&(1-z^2)Z''(z)+\{2a-2b-(2a+2b+2)z\}Z'(z)\nonumber \\&&\qquad +\{k_1-(a+b)(a+b+1)\}Z(z)=0,\label{pr5}
\end{eqnarray}
where $2a=\nu_1-2\nu_2$, $2b=\nu_1$.
Comparing (\ref{pr5}) with the Jacobi differential equation 
\begin{equation}
(1-x^{2})y''+\{\beta_1-\alpha_1-(\alpha_1+\beta_1+2)x\}y'+\lambda(\lambda+\alpha_1+\beta_1+1)y=0,\label{Jd1}
\end{equation}
we obtain the  separation constant 
\begin{equation}
k_1=(l-\nu_2)(l-\nu_2+1),\label{pr2} 
\end{equation}
where $l=\lambda+\nu_1$. 
Hence the solutions of (\ref{pr5}) are given in terms of the Jacobi polynomials as
\begin{eqnarray}
\Theta(\theta)&\equiv &\Theta_{l \nu_1}(\theta; \nu_{1}, \nu_{2})
= F_{l \nu_1}(\nu_{1}, \nu_{2})(1+\cos\theta)^{\frac{(\nu_1-2\nu_2)}{2}}(1-\cos\theta)^{\frac{\nu_1}{2}}\nonumber\\&&\quad\times P^{(\nu_1, \nu_1-2\nu_2)}_{l-\nu_1}(\cos\theta),\label{jp1}
\end{eqnarray}
where $P^{(\alpha, \beta)}_{\lambda}$ denotes a Jacobi polynomial \cite{Mag1}, $F_{l \nu_1}(\nu_{1}, \nu_{2})$ is the normalized constant and $l\in \mathbb{N}$.

The radial equation (\ref{kf5}) can be converted, by setting  
 $z=\varepsilon r^2$, $R(z)=z^{\frac{1}{2}( l-\nu_2)} e^{-\frac{z}{2}}R_1(z)$ and $\varepsilon^2=\frac{c_0}{2}-2aE+d\nu_2^2$, to
\begin{eqnarray}
&&z\frac{d^2R_1(z)}{dz^2}+\left\{(l-\nu_2+\frac{3}{2})-z\right\}\frac{dR_1(z)}{dz}-\left\{\frac{1}{2}(l-\nu_2+\frac{3}{2})+\frac{E'}{4\varepsilon}\right\}R_1(z)=0.\nonumber\\&&\label{an11}
\end{eqnarray}
Set  
\begin{eqnarray}
 n=\frac{\nu_{2}}{2}-\frac{E'}{4\varepsilon}-\frac{l}{2}-\frac{3}{4}.\label{an12}
\end{eqnarray}
Then (\ref{an11}) can be identified with the Laguerre differential equation. Hence we can write the solution of (\ref{kf5}) in terms of the confluent hypergeometric polynomial as  
\begin{eqnarray}
&R(r)&\equiv R_{nl}(r;\nu_{1}, \nu_{2})=F_{nl}(\nu_{1},\nu_{2})(\varepsilon r^2)^{\frac{1}{2}(l-\nu_2)} e^{\frac{-\varepsilon r^2}{2}}\nonumber\\&&
\quad \times {}_1 F_1(-n, l-\nu_2+\frac{3}{2}; \varepsilon r^2),\label{an14}
\end{eqnarray}
where $F_{nl}(\nu_{1},\nu_{2})$ is the normalized constant.
In order to have a discrete spectrum the parameter $n$ needs to be positive integer. From (\ref{an12}) 
\begin{equation}
\varepsilon=\frac{-E'}{4n+2l-2\nu_{2}+3}\label{en1}
\end{equation} and hence the energy spectrum of the system (\ref{kf1}) is given by
\begin{equation}
\frac{2bE-c_1\nu_2^2-c_4 }{\sqrt{\frac{c_0}{2}-2a E+d\nu_2^2}}=4n+2l-2\nu_2+3,\quad n=1, 2, 3,\dots\label{en1}
\end{equation}

\subsection{Ladder operators and recurrence approach}
The solutions of the eigenfunction for the equation $H'\Psi=E'\Psi$ of the form $\Psi(r,\theta,\phi,\psi)=\psi^{l-\nu_2+\frac{1}{2}}_{n}\Theta^{( \nu_1,\nu_1-2\nu_2)}_{l-\nu_1} e^{i(\nu_1\phi+\nu_2\psi)}$ are given by
\begin{eqnarray}
&&\psi^{l-\nu_2+\frac{1}{2}}_{n}=e^{\frac{-\varepsilon r^2}{2}}r^{l-\nu_2} L^{l-\nu_2+\frac{1}{2}}_n
(\varepsilon r^2),
\\&&
\Theta^{(\nu_1, \nu_1-2\nu_2)}_{l-\nu_1}=\sin^{\nu_1}{\frac{\theta}{2}}\cos^{\nu_1-2\nu_2}{\frac{\theta}{2}}P^{( \nu_1, \nu_1-2\nu_2)}_{l-\nu_1}(\cos\theta),
\end{eqnarray} 
where $L^\alpha_n$ is the $n$th order Laguerre polynomial, $P^{(\beta,\gamma)}_\lambda$ is the Jacobi polynomial \cite{Mag1},  $n=\frac{\nu_2}{2}-\frac{E'}{4\varepsilon}-\frac{3}{4}-\frac{l}{2}$ and $\varepsilon^2=\frac{c_0}{2}-2aE+d\nu_2^2$. The energy eigenvalues of the equation $H'\Psi=E'\Psi$ is
\begin{eqnarray}
E'=-\varepsilon(4n+2l-2\nu_2+3).
\end{eqnarray}
Let us now construct the ladder operators based on radial part of the separated solutions using differential identities for Laguerre functions \cite{Mag1}
\begin{eqnarray}
&&K^{+}_{l-\nu_2+\frac{1}{2},n}=\frac{1}{r}(B-Q+1)\partial_{r}-\frac{H'}{2}+\frac{1}{r^2}(B-Q+1)(B-Q-\frac{1}{2}),\nonumber
\\&&
\\&&
K^{-}_{l-\nu_2+\frac{1}{2},n}=-\frac{1}{r}(B-Q-1)\partial_{r}-\frac{H'}{2}-\frac{1}{r^2}(B-Q-1)(B-Q+\frac{1}{2})
\nonumber
\\&&
\end{eqnarray}
and the shift operators based on the angular functions using Jacobi function identities \cite{Mag1}
\begin{eqnarray}
&J^{+}_{l-\nu_1} &=-2(B-Q+\frac{1}{2})\sin\theta \partial_\theta-2(B-Q+\frac{1}{2})^2\cos\theta -2Q(L_3-Q),\nonumber\\&&
\\
&J^{-}_{l-\nu_1} &=2(B-Q-\frac{1}{2})\sin\theta \partial_\theta-2(B-Q-\frac{1}{2})^2\cos\theta -2Q(L_3-Q).\nonumber\\&&
\end{eqnarray}
Here $B=\sqrt{\textbf{L}^2+\frac{1}{4}}$ as in section 2.
The action of the operators on the corresponding wave functions provide the following recurrence formulas
\begin{eqnarray}
&&K^{+}_{l-\nu_2+\frac{1}{2},n}\psi^{l-\nu_2+\frac{1}{2}}_{n}=-2\varepsilon^2 \psi^{l-\nu_2+\frac{5}{2}}_{n-1},
\\&&
K^{-}_{l-\nu_2+\frac{1}{2},n}\psi^{l-\nu_2+\frac{1}{2}}_{n}=- 2(n+1)(n+l-\nu_2+\frac{1}{2}) \psi^{l-\nu_2-\frac{3}{2}}_{n+1},
\\
&&J^{+}_{l-\nu_1} \Theta^{(\nu_1, \nu_1-2\nu_2)}_{l-\nu_1}=-2(l-\nu_1+1)(l-\nu_1-2\nu_2+1)\Theta^{(\nu_1, \nu_1-2\nu_2)}_{l-\nu_1+1},
\\&&
J^{-}_{l-\nu_1} \Theta^{(\nu_1, \nu_1-2\nu_2)}_{l-\nu_1}=-2l(l-2\nu_2)\Theta^{(\nu_1, \nu_1-2\nu_2)}_{l-\nu_1-1}.
\end{eqnarray}

\subsection{Integrals of motion, algebra structure and spectrum}
Let us now consider the suitable operators 
\begin{eqnarray}
D_1=K^{+}_{l-\nu_2+\frac{1}{2},n}J^{+}_{l-\nu_1+1}J^{+}_{l-\nu_1} B, \quad D_2=BJ^{-}_{l-\nu_1}J^{-}_{l-\nu_1-1}K^{-}_{l-\nu_2+\frac{1}{2},n}.
\end{eqnarray}
The explicitly action of the operators $D_i, i=1, 2$ on the wave functions is given by
\begin{eqnarray}
D_1\Psi(r,\theta,\phi,\psi)&=&-8\varepsilon^2l(l+1)(l-\nu_1+1)(l-\nu_1-2\nu_2+1)(l-\nu_1+2)\nonumber\\&&\times(l-\nu_1-2\nu_2+2) \psi^{l-\nu_2+\frac{5}{2}}_{n-1}\Theta^{(\nu_1, \nu_1-2\nu_2)}_{l-\nu_1+2}e^{i(\nu_1\phi+\nu_2\psi)},
\\
D_2\Psi(r,\theta,\phi,\psi)&=&-8l(l-1)(l-2\nu_2)(l-2\nu_2-1)(l-\frac{3}{2})(n+1)\nonumber\\&&\times(n+l-\nu_2+\frac{1}{2}) \psi^{l-\nu_2-\frac{3}{2}}_{n+1}\Theta^{(\nu_1, \nu_1-2\nu_2)}_{l-\nu_1-2}e^{i(\nu_1\phi+\nu_2\psi)}.
\end{eqnarray}
We can also obtain the action of the operators $D_1D_2$ and $D_2D_1$ on the wave functions. It follows in the operator from construction they form an algebraically independent set of differential operators and there has a common feature of superintegrable systems to close polynomially symmetry algebra. Direct computation shows that they form higher order polynomial algebra 
\begin{eqnarray}
&&[D_1, H']=0=[D_2, H'],\label{ff1}
\\&& [B, D_1]= 2D_1, \quad [B, D_2]=-2D_2,\label{ff2}
\end{eqnarray}
\begin{eqnarray}
&D_1D_2&=\frac{B-2}{256}(2B-5)(2B-3)^2(2B-2L_3-4Q-1)\nonumber\\&&\times (2B-2L_3-1)(2B-4Q-3)(2B-4Q-1)\nonumber\\&&\times(2B-2L_3-4Q-3)(2B-1)(2B-2L_3-3)\nonumber\\&&\times [H'-2\varepsilon(B-Q-1)][H'+2\varepsilon(B-Q-1)],\label{ff3}
\end{eqnarray}
\begin{eqnarray}
&D_2D_1&=\frac{B}{256}(2B-1)(2B+1)^2 (2B-2L_3-4Q+3)\nonumber\\&&\times (2B-2L_3+3)(2B-4Q+1)(2B-4Q+3)\nonumber\\&&\times(2B-2L_3-4Q+1)(2B+3)(2B-2L_3+1)\nonumber\\&&\times [H'-2\varepsilon(B-Q+1)][H'+2\varepsilon(B-Q+1)].\label{ff4}
\end{eqnarray}
We rewrite ((\ref{ff1})-(\ref{ff4})) in the form of deformed oscillator algebra (\ref{kpfh}) by letting $\aleph=\frac{B}{2}$,  $b^{\dagger}=D_1$ and $b=D_2$. We then obtain structure function
\begin{eqnarray}
&\Phi(x;u,H')&=\frac{(2x+u-2)}{256}[2(2x+u)-3]^2[2(2x+u)-2L_3-1]\nonumber\\&&\times[2(2x+u)-1] [2(2x+u)-2L_3-3][2(2x+u)-5]\nonumber\\&&\times[2(2x+u)-2L_3-4Q-3][H'-2\varepsilon\{(2x+u)-Q-1\}]\nonumber\\&&\times[2(2x+u)-4Q-1][H'+2\varepsilon\{(2x+u)-Q-1\}]\nonumber\\&&\times [2(2x+u)-4Q-3][2(2x+u)-2L_3-4Q-1] ,
\end{eqnarray}
where $u$ is an arbitrary constant to be determined. We should impose the following constraints on the structure function in order to obtain a finite dimensional unirreps, 
\begin{equation}
\Phi(p+1; u,E')=0,\quad \Phi(0;u,E')=0,\quad \Phi(x)>0,\quad \forall x>0,\label{pro2}
\end{equation}
where $p$ is a positive integer. These constraints give $(p+1)$-dimensional unitary representations and their solutions give the energy $E'$ and the arbitrary constant $u$. We have all the possible energy spectra and structure functions as 
\begin{eqnarray}
u=\frac{\varepsilon_1 E'+2\varepsilon(1+\nu_2)}{2\varepsilon}, \quad E'=-\varepsilon(4p-2\nu_1+2\varepsilon_2\nu_2+3)
\end{eqnarray}
\begin{eqnarray}
&\Phi(x)&=2\varepsilon^2[2x-2p+(\varepsilon_2-1)\nu_1-2][2x-2p+(\varepsilon_2-1)\nu_1-2\nu_2-2]\nonumber\\&&\times[4x-(4p+2\varepsilon_2\nu_1+2\nu_2+3)(1+\varepsilon_1)][2x-2p+\varepsilon_2\nu_1-2\nu_2-1]\nonumber\\&&\times[4x-(4p+2\varepsilon_2\nu_1+2\nu_2+3)(1-\varepsilon_1)][2x-2p+\varepsilon_2\nu_1-2\nu_2-2]\nonumber\\&&\times[2x-2p+\varepsilon_2-2][2x-2p+\varepsilon_2\nu_1-3][2x-2p+\varepsilon_2\nu_1-1]\nonumber\\&&\times[4p-2x+2\varepsilon_2\nu_1+5][2x-2p+\varepsilon_2\nu_1-2]^2\nonumber\\&&\times[2x-2p+(\varepsilon_2-1)\nu_1-2\nu_2-1],
\end{eqnarray}
where $\varepsilon_1=\pm 1$, $\varepsilon_2=\pm 1$. The coupling constant metamorphosis provides $E'\leftrightarrow c_4+c_1\nu_2^2-2bE$ and $\varepsilon^2\leftrightarrow \frac{c_0}{2}-2aE+d\nu_2^2$. Hence we have the energy  of the original Hamiltonian
\begin{eqnarray}
\frac{2bE-c_1\nu_2^2-c_4 }{\sqrt{\frac{c_0}{2}-2a E+d\nu_2^2}}=4p-2\nu_1+2\varepsilon_2 \nu_2+3.\label{en2}
\end{eqnarray}
Making the identifications $p=n$, $-\nu_1=l$, $\varepsilon_1=1$, $\varepsilon_2=1$, then (\ref{en2}) coincides with the physical spectra (\ref{en1}).

\section{Conclusion}
One of the main results of this paper is the construction via recurrence method of the higher order integrals of motion and higher rank polynomial algebra for the MIC-harmonic oscillator systems with monopole interactions in both flat space and curved Taub-NUT space. The method is systematic and well constructed based on wave functions of the systems. To our knowledge this is the first application of the recurrence approach in superintegrable monopole systems.

Let us point out that superintegrable systems with monopole interactions and their polynomial algebras are largely unexplored area  \cite{Mar10}. It is interesting to generalize the results to systems with non-abelian monopole interactions. Research in superintegrable system with Yang-Coulomb monopole is underway and results will be presented elsewhere.

{\bf Acknowledgements:}
The research of FH was supported by International Postgraduate Research Scholarship and Australian Postgraduate Award. IM was supported by the Australian Research Council through a Discovery Early Career Researcher Award DE 130101067. YZZ was partially supported by the Australian Research Council, Discovery Project DP 140101492. He would like to thank the Institute of Theoretical Physics, Chinese Academy of Sciences, for hospitality and support.

\end{document}